\input harvmac
\input epsf

\font\tenams=msam10 \font\sevenams=msam7 
\newfam\amsfam 
\textfont\amsfam=\tenams
\scriptfont\amsfam=\sevenams  

\def\hexnumber#1{\ifcase#1 0\or 1\or 2\or 3\or 4\or 5\or 6\or 7\or
8\or 9\or A\or B\or C\or D\or E\or F\fi}

\edef\theamsfam{\hexnumber\amsfam}  

\mathchardef\gtrsim="3\theamsfam 26 
\mathchardef\lesssim="3\theamsfam 2E 

\ifx\epsfbox\UnDeFiNeD\message{(NO epsf.tex, FIGURES WILL BE
IGNORED)}
\def\figin#1{\vskip2in}
\else\message{(FIGURES WILL BE INCLUDED)}\def\figin#1{#1}\fi
\def\ifig#1#2#3{\xdef#1{fig.~\the\figno}
\midinsert{\centerline{\figin{#3}}%
\smallskip\centerline{\vbox{\baselineskip12pt
\advance\hsize by -1truein\noindent{\bf Fig.~\the\figno:} #2}}
\bigskip}\endinsert\global\advance\figno by1}
\noblackbox

\def\at{{\tilde a}}
\def\Xt{{\tilde X}}
\def\omegat{{\tilde \omega}}
\def\xit{{\tilde \xi}}
\def\Ft{{\tilde F}}
\def\qt{{\tilde q}}

%
%

%

\lref\douglas{M.~R.~Douglas and S.~H.~Shenker,
``Dynamics of SU(N) supersymmetric gauge theory,''
Nucl.\ Phys.\ B {\bf 447}, 271 (1995)
[arXiv:hep-th/9503163].}
\lref\argyres{P.~C.~Argyres and A.~E.~Faraggi,
``The vacuum structure and spectrum of N=2 supersymmetric SU(n) gauge theory,''
Phys.\ Rev.\ Lett.\  {\bf 74}, 3931 (1995)
[arXiv:hep-th/9411057].}
\lref\argyrdoug{P.~C.~Argyres and M.~R.~Douglas,
``New phenomena in SU(3) supersymmetric gauge theory,''
Nucl.\ Phys.\ B {\bf 448}, 93 (1995)
[arXiv:hep-th/9505062].}
\lref\theisen{A.~Klemm, W.~Lerche, S.~Yankielowicz and S.~Theisen,
``Simple singularities and N=2 supersymmetric Yang-Mills theory,''
Phys.\ Lett.\ B {\bf 344}, 169 (1995)
[arXiv:hep-th/9411048].}
\lref\dhoker{E.~D'Hoker and D.~H.~Phong,
``Lectures on supersymmetric Yang-Mills theory and integrable systems,''
arXiv:hep-th/9912271.}
\lref\ferrari{F.~Ferrari,
``The large N limit of N = 2 super Yang-Mills, fractional instantons and  infrared divergences,''
Nucl.\ Phys.\ B {\bf 612}, 151 (2001)
[arXiv:hep-th/0106192].}
\lref\fstring{F.~Ferrari,
``Exact amplitudes in four dimensional non-critical string theories,''
Nucl.\ Phys.\ B {\bf 617}, 348 (2001)
[arXiv:hep-th/0107096].}
\lref\constable{N.~R.~Constable,
``The entropy of 4D black holes and the enhancon,''
Phys.\ Rev.\ D {\bf 64}, 104004 (2001)
[arXiv:hep-th/0106038].}
\lref\jmpr{C.~V.~Johnson, R.~C.~Myers, A.~W.~Peet and S.~F.~Ross,
``The enhancon and the consistency of excision,''
Phys.\ Rev.\ D {\bf 64}, 106001 (2001)
[arXiv:hep-th/0105077].}
\lref\johnson{C.~V.~Johnson,
``Enhancons, fuzzy spheres and multi-monopoles,''
Phys.\ Rev.\ D {\bf 63}, 065004 (2001)
[arXiv:hep-th/0004068].}
\lref\johnsontwo{C.~V.~Johnson,
``The enhancon, multimonopoles and fuzzy geometry,''
Int.\ J.\ Mod.\ Phys.\ A {\bf 16}, 990 (2001)
[arXiv:hep-th/0011008].}
\lref\dhkp{E.~D'Hoker, I.~M.~Krichever and D.~H.~Phong,
``The effective prepotential of N = 2 supersymmetric SU(N(c)) gauge  theories,''
Nucl.\ Phys.\ B {\bf 489}, 179 (1997)
[arXiv:hep-th/9609041].}
\lref\jpp{C.~V.~Johnson, A.~W.~Peet and J.~Polchinski,
``Gauge theory and the excision of repulson singularities,''
Phys.\ Rev.\ D {\bf 61}, 086001 (2000)
[arXiv:hep-th/9911161].}
\lref\peet{A.~W.~Peet,
``More on singularity resolution,''
arXiv:hep-th/0106148.}
\lref\johnmy{C.~V.~Johnson and R.~C.~Myers,
``The enhancon, black holes, and the second law,''
Phys.\ Rev.\ D {\bf 64}, 106002 (2001)
[arXiv:hep-th/0105159].}
\lref\seiberg{N.~Seiberg and E.~Witten,
``Electric - magnetic duality, monopole condensation, and confinement in N=2 supersymmetric Yang-Mills theory,''
Nucl.\ Phys.\ B {\bf 426}, 19 (1994)
[Erratum-ibid.\ B {\bf 430}, 485 (1994)]
[arXiv:hep-th/9407087].}
\lref\brandeis{J.~M.~Isidro, A.~Mukherjee, J.~P.~Nunes and H.~J.~Schnitzer,
``A new derivation of the Picard-Fuchs equations for effective N = 2  super Yang-Mills theories,''
Nucl.\ Phys.\ B {\bf 492}, 647 (1997)
[arXiv:hep-th/9609116].}

\vskip 1cm

 \Title{ \vbox{\baselineskip12pt\hbox{  BROWN HET-1306 }}}
 {\vbox{
\centerline{  Moduli Space Metric of ${\cal N}=2$ Supersymmetric }
\vskip.08in
\centerline{ $SU(N)$ Gauge Theory  and the Enhancon   } }}

\centerline{$\quad$ {Gian Luigi Alberghi, Steven Corley, and David A. Lowe   } }
\smallskip
\centerline{{\sl Department of Physics}}
\centerline{{\sl Brown  University}}
\centerline{{\sl Providence, RI 02912 }}
\smallskip
\centerline{{\tt gigi@het.brown.edu }}
\centerline{{\tt   scorley@het.brown.edu }}
\centerline{{ \tt lowe@het.brown.edu }}

\vskip .3in 

We compute the moduli space metric of $SU(N)$ Yang-Mills theory with 
${\cal N} =2$ supersymmetry in the vicinity of the point where the classical
moduli vanish. This gauge theory may be realized as a set of $N$
D7-branes wrapping a $K3$ surface, near the enhancon locus.
The moduli space metric determines the low-energy worldvolume dynamics of
the D7 branes near this point, including stringy corrections. Non-abelian
gauge symmetry is not restored on the worldvolume at the enhancon
point, but rather the gauge group remains $U(1)^{N-1}$ and light
electric and magnetically charged particles coexist. We also study
the moduli space metric for a single probe brane in the background of
$N-1$ branes near the enhancon point. We find quantum corrections to
the supergravity probe metric that are not suppressed at
large separations, but are down by $1/N$ factors, 
due to the response of the $N-1$ enhancon branes to
the probe. 
A singularity appears before the
probe reaches the enhancon point where a dyon becomes
massless. We compute the masses of W-bosons and monopoles in
a large $N$ limit near this critical point.


\Date{ March 2002 }

\newsec{ Introduction } 

Some time ago, Argyres and Faraggi \argyres, and Klemm, Lerche, Theisen and
Yankielowicz \theisen\ obtained the exact solution of the low-energy effective
action of ${\cal N}=2$
supersymmetric $SU(N)$ gauge theory, generalizing the work of Seiberg
and Witten for $SU(2)$ \seiberg. The answer is written as a prepotential for
the ${\cal N}=2$ theory, expressed as the period matrix of a
certain hyperelliptic Riemann surface. The period matrix becomes the
metric for the $N-1$ dimensional moduli space of this gauge theory.
This period matrix was
evaluated in an instanton expansion, in the limit of well-separated
moduli, by D'Hoker, Kirchever and Phong \dhkp. Picard-Fuchs equations
were set up by \brandeis. Many of these results are described in the
review article \dhoker.

The large $N$ limit of this theory was considered by Douglas and
Shenker \douglas. This limit is rather subtle, because if the moduli
have finite separations as $N\to \infty$ the instanton corrections
vanish exponentially fast. To obtain a non-trivial limit, one must
consider points on moduli space where separations vanish as $1/N$. 
Douglas and Shenker focused on the case where the moduli eigenvalues
are  located on the real axis. This is the case that is smoothly
connected to vacua of the ${\cal N}=1$ gauge theory with massive
adjoint matter. In this paper we will generalize the analysis of
\douglas, to the case when the moduli sit near a circle (or circles)
in the complex plane, as well as to the case where a pair of
eigenvalues are separated from such a circle. This limit has also been 
studied in \refs{\ferrari, \fstring}.

One of the main motivations for studying this problem is given by
recent attempts to understand the resolution of certain naked
singularities in string theory. For the case of D-branes wrapping a
$K3$ surface, it was argued \jpp\ that the timelike repulson singularity
present in the naive continuation of the geometry, should be replaced
by a shell of D-brane source sitting outside this point -- the
so-called enhancon locus. Inside this locus, it was argued the
geometry should be flat. This picture received further support from a
detailed analysis of the junction conditions \jmpr, and from black
hole entropy considerations \refs{\johnmy,\constable}. 

A detailed understanding of this phenomena requires a full string
theory treatment of the problem. From the point of view of the
supergravity analysis, the D-brane source boundary conditions are put
in by hand. In this paper we present the nonperturbative moduli space
metric that describes the low-energy worldvolume dynamics of D7-branes
wrapping $K3$, near the enhancon locus. This is an important step
toward a better string theory understanding of the enhancon
phenomenon.

Our results support the picture advocated in \jpp. A wrapped D7-brane
probe approaching $N-1$ branes at the enhancon point, smoothly melts into the
other branes, and then emerges from the other side. The probe brane
does not see the interior of the geometry at all. The enhancon locus
is simply a non-singular point from the viewpoint of the
probe. Interestingly, we find nontrivial quantum corrections to the
probe metric that do not fall off at large separations, however these
appear at subleading order in $1/N$.

The plan of this paper is as follows: in section 2 we briefly review
the solution of $SU(N)$ ${\cal N}=2$ supersymmetric gauge theory of
\refs{\argyres,\theisen} to fix notation; in section 3 we compute the
moduli space metric along a one-complex parameter slice of the space
in two different situations: for a pair of enhancon-like shells, and
for a probe brane in an enhancon background. An analytic form of the
metric along this slice is presented for all values of $N$. The metric
includes nontrivial quantum corrections. For the
case of the probe brane, the moduli space metric has a singularity as
the probe starts to get close to the enhancon shell, when two
eigenvalues collide and a dyon becomes massless. 
We obtain analytic expressions for the masses of dyons and
monopoles in the large $N$ limit, near this critical point. In section
4 we comment on the implications of these results for the resolution
of spacetime singularities in string theory, and discuss future prospects.

\newsec{ Review and Notation  }

Recall that the solution to the low-energy effective action of ${\cal
N}=2$ supersymmetric $SU(N)$ Yang-Mills theory is given in terms of an
auxiliary Riemann surface ${\cal C}$, that consists of a genus
$N-1$ hyperelliptic curve ${\cal C}$ defined by
\eqn\curve{Y^2 = \prod_{j=1}^{N} (X - \phi_j)^2 - \Lambda^{2N}
= (P(X))^2 - \Lambda^{2N}}
along with the 1-form
\eqn\swform{ \lambda = {X dP \over 2 \pi i Y}}
whose $\phi_j$ derivatives are holomorphic. The $\phi_i$ denote the
classical expectation values of the gauge theory moduli. For
SU(N) gauge theory the moduli space coordinates must
satisfy the tracelessness condition
\eqn\traceless{\sum_{j=1}^{N} \phi_j = 0.}  
The periods
$a_j$ and $a_{Dk}$ are then given by the integrals
\eqn\periods{a_j = \oint_{\beta_j} \lambda, \hskip6pt
a_{D{m}} = \oint_{\alpha_m} \lambda} for some choice
of cycles $\alpha_m$ and $\beta_k$ where only $N-1$ of each of
these sets of cycles is independent. 
 
To construct the moduli space metric one requires 
derivatives of the periods with respect to the moduli
space coordinates $\phi_k$.  If the periods \periods\
can be evaluated as functions of the coordinates
$\phi_j$ then this is easy to compute.  Finding the
periods at a generic point in moduli space however
is difficult and it is easier instead to compute
the derivatives at a fixed point in moduli space
directly by noting that
\eqn\dlamdphi{{\partial \lambda \over \partial \phi_k} =
- {1 \over i 2 \pi Y} {\partial P \over \partial \phi_k} dX
+ d\left({X \over Y} {\partial P \over \partial \phi_k} \right). }
The term inside the total derivative is periodic around a closed
contour, so
the $\phi$-derivatives of the period integrals can be
evaluated using only the first term of \dlamdphi.

Using this information one can then compute the period
matrix $\tau_{mk}$ via the prescription
\eqn\tausw{\tau_{mk} = {\partial a_{Dm} \over \partial a_k}
= \sum_{l=1}^{N-1} {d a_{Dm} \over d \phi_l}
 \left( {d a \over d \phi}\right)^{-1}_{lk} }
where the derivatives on the far right-hand-side are
total derivatives.  That is, we have removed the $\phi_N$
dependence using the tracelessness property \traceless\ so that
\eqn\totalderiv{{d a_{Dm} \over d \phi_l} =
{\partial a_{Dm} \over \partial \phi_l} -
{\partial a_{Dm} \over \partial \phi_N}.}
It is important to note however that this prescription has
an apparent singularity if $\phi_j = \phi_k$ for any indices
$j \neq k$.  The singularity can be easily seen from \dlamdphi\ by noting
that $\partial a_l/\partial \phi_j = \partial a_l/\partial \phi_k$
when evaluated at the point $\phi_j = \phi_k$.  The 
$(N-1) \times (N-1)$ matrix $[\partial a/\partial \phi ]$ is then not
invertible.  If the branch points of the curve \curve\ are
all distinct, then we do not expect this point in moduli
space to be a real singularity, but rather just a coordinate
singularity.  In the cases that we discuss below we will
indeed see that this is true.

\newsec{Moduli Space Metric} 

It is rather difficult to obtain explicit expressions for the 
metric throughout the full moduli space. To simply matters
we will focus 
on two distinct one complex parameter slices through the full
$N-1$-dimensional moduli space, each of which contains the enhancon
point (all $\phi_i = 0$). 
 
\subsec{Two enhancon shells} 

The first slice we consider retains a $Z_N$ rotation symmetry in the
curve ${\cal C}$, and
corresponds to a pair of concentric enhancon shells, which coalesce at
the enhancon point. The corresponding brane configurations have been
analyzed from the supergravity viewpoint in \refs{\johnmy,\jmpr,\constable}.
We parametrized the moduli space coordinates in
terms of a complex parameter $v$ as
\eqn\doublecoords{\phi_j = v ~ e^{i 2 \pi j/N}}
where $1 \leq j \leq N$.  Note that this choice of 
moduli space coordinates does satisfy the tracelessness condition
\traceless.  The enhancon point sits at $v=0$, which leads to a 
coordinate singularity in the period matrix
\tausw\ as discussed above. 
One can get around this problem
as we shall see by computing the period matrix
for nonzero $v$ and then taking the limit $v \rightarrow 0$,
in which case a nonsingular answer is obtained.  For
nonzero $v$ this subspace corresponds to two separate enhancon
shells.

\ifig\fone{Positions of branch cuts, and choice of 1-cycles for double
enhancon configuration.}{\epsfysize3.0in\epsfbox{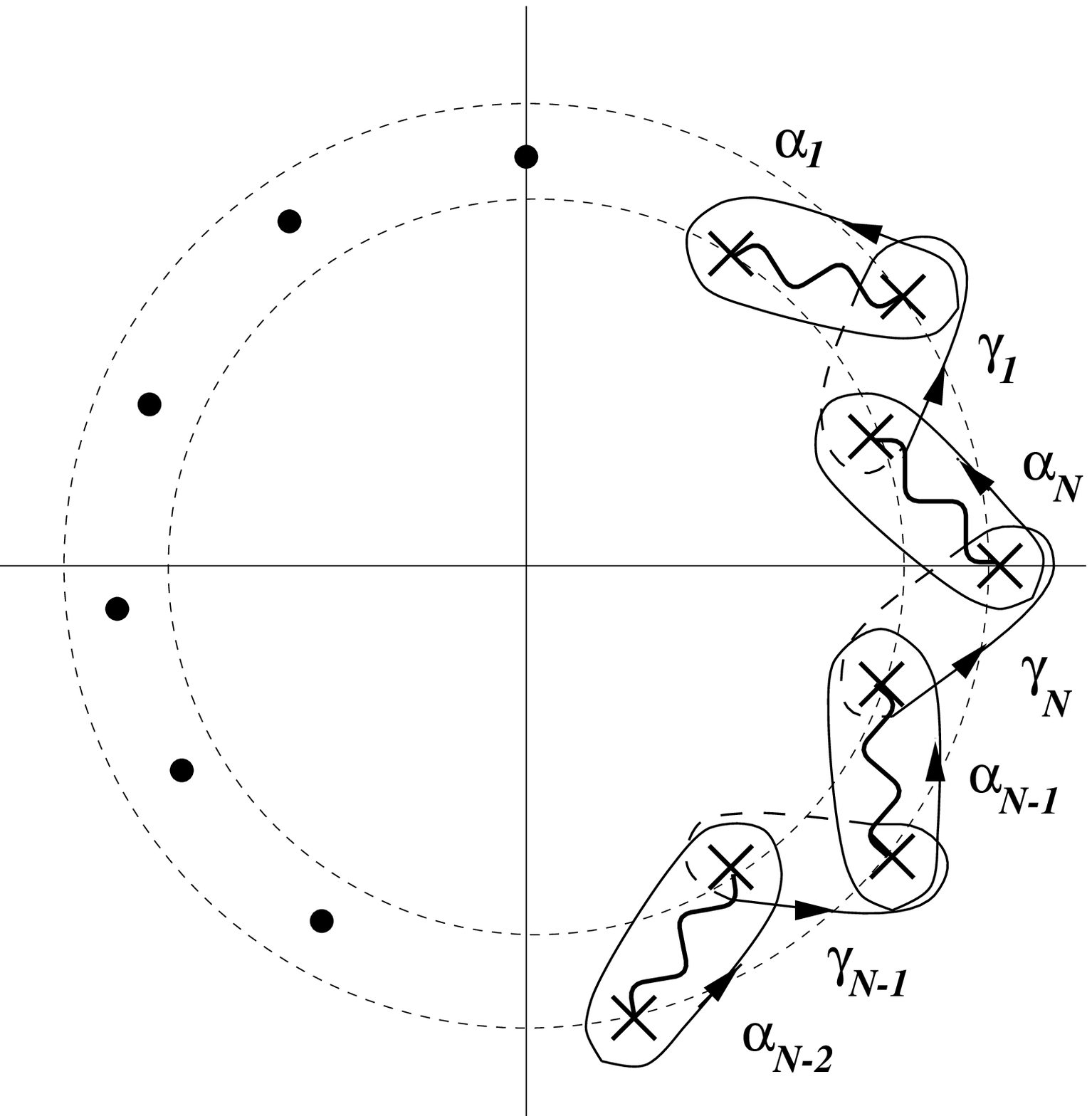}}

To see this in more detail we must solve for the branch points
of the curve given by $Y = 0$.  For the simple subspace in 
moduli space that we have taken these can be found exactly.
The point is that the product in \curve\ simplifies to
\eqn\prodone{\prod_{j=1}^{N} (X - \phi_j) = X^N - v^N,}
which follows easily from the identity
\eqn\sumident{\sum_{j=1}^{N} e^{i 2 \pi j k/N}
=N \delta_{k,0~ {\rm mod} ~N}~.}
The branch points are then found to be
\eqn\branchdouble{\eqalign{X_{2k} & = (\Lambda^N + v^N)^{1/N} 
e^{i 2 \pi k/N}, \cr
X_{2k+1} & = (\Lambda^N - v^N)^{1/N} 
e^{i 2 \pi (k+ 1/2)/N}}}
for $1 \leq k \leq N$.
For real $v$ we get the structure shown in \fone\ with
a pair of concentric circles of branch points.  We have
also shown some of the cycles $\alpha_m$ and $\gamma_j$.
We take the $\alpha_m$ cycle to enclose the $X_{2m}$ and
$X_{2m+1}$ branch points, with $1 \leq m < N$. The $\gamma_j$ cycle
encloses the $X_{2j-1}$ and $X_{2j}$ branch points, with $1\leq j \leq
N$.
The $\beta_n$ cycles are then given in terms of the
$\gamma$ cycles by
\eqn\betacycles{\beta_n = \sum_{j=1}^{n} \gamma_j}
and we take $1\leq n < N$.
The $\alpha$ and $\beta$ cycles comprise a canonical
basis of homology 1-cycles.

The holomorphic 1-forms \dlamdphi\ on this slice of moduli space
are given by
\eqn\dldpdouble{{\partial \lambda \over \partial \phi_k}
= - {1 \over i 2 \pi}{(X^N - v^N) ~ dX \over (X - v ~ e^{i 2 \pi k/N})
\sqrt{(X^N - v^N)^2 - \Lambda^{2N}}}.}
While one could try to compute the period matrix using this
basis of holomorphic 1-forms,
it turns out that it is not the most convenient. The computation of
the period matrix is simpler if we change basis to
\eqn\newdldpdouble{\eqalign{{\partial \lambda \over \partial t_n}
& = \sum_{k=1}^{N} {\partial \lambda \over \partial \phi_k}
e^{i 2 \pi k n/N} \cr
& = \sum_{k=1}^{N-1} \left({\partial \lambda \over \partial \phi_k}
- {\partial \lambda \over \partial \phi_N}\right)
e^{i 2 \pi k n/N} \cr
& = - {1 \over i 2 \pi N} \left({v \over X}\right)^{N-n} {d ~X^N \over
\sqrt{(X^N - v^N)^2 - \Lambda^{2N}}}}}
where the second line follows from the identity \sumident.

To compute the period matrix we now compute the derivatives of
the periods
\eqn\periodderivs{{\partial a_j \over \partial t_n} = \oint_{\beta_j} 
{\partial \lambda \over \partial t_n}, \qquad
{\partial a_{D{m}} \over \partial t_n} = \oint_{\alpha_m} {\partial \lambda \over
\partial t_n}.}
For the $a_D$-periods we make the change of variables
\eqn\covaD{
X = (\Lambda^N e^{i \varphi} + v^N)^{1/N} e^{i 2 \pi j/N}}
which reduces \newdldpdouble\ to
\eqn\newtwodldp{{\partial \lambda \over \partial t_n} = - {v^{N-n} \over 2 \pi N}
e^{i 2 \pi j n/N} {1 \over (\Lambda^N e^{i \varphi} + v^N)^{1-n/N}}
{e^{i \varphi} d\varphi \over \sqrt{e^{i 2 \varphi} -1}}.}
Defining 
\eqn\fdefn{f(v^N,n/N) \equiv - {1 \over \pi N}
\int_{0}^{\pi} ~ d\varphi ~ {1 \over (\Lambda^N e^{i \varphi} + v^N)^{1-n/N}}
{e^{i \varphi} \over \sqrt{e^{i 2 \varphi} -1}}}
we then have
\eqn\daDdt{{\partial a_{Dm} \over \partial t_n} = v^{N-n} f(v^N,n/N)
e^{i 2 \pi m n/N}.}
Similarly for the $a$-periods we find
\eqn\datildedt{{\partial {\tilde a}_{k} \over \partial t_n} = v^{N-n} 
f(-v^N,n/N)
e^{i 2 \pi (k-1/2) n/N}.}
The tilde notation here denotes a period evaluated along a
$\gamma$ contour.  To obtain the $a$-period we must sum as
in \betacycles, or in matrix notation  
\eqn\atotildea{{\partial a_{j} \over \partial t_n} = \sum_{k=1}^{N-1} q_{jk}
{\partial {\tilde a}_{k} \over \partial t_n}}
where the matrix entries of $q_{jk}$ are given by ones for
$j \leq k$ and zeroes everywhere else. 

What makes the coordinate transformation in \newdldpdouble\ so
convenient is that \daDdt\ and \datildedt\ depend on the $a_D$
and ${\tilde a}$ indices respectively only through a phase factor.
Consequently either matrix is easy to invert provided
that $f(v^N,n/N)$ does not vanish for any $n$ and $v$.
Specifically the $(N-1) \times (N-1)$ matrix 
\eqn\mjn{m_{jn} \equiv e^{i 2 \pi j n/N}}
has the inverse
\eqn\minverse{(m^{-1})_{nk} = {1 \over N}(e^{-i 2 \pi n k/N} -1)}
from which it follows that
\eqn\inverses{\eqalign{\left({\partial a_{D} \over \partial t}\right)^{-1}_{np} & = 
{1 \over v^{N-n} f(v^N,n/N)} (m^{-1})_{np}, \cr
\left({\partial {\tilde a} \over \partial t}\right)^{-1}_{nj} & = 
{e^{i \pi n/N} \over v^{N-n} f(-v^N,n/N)} (m^{-1})_{nj}.
}}
Using \daDdt\ and \inverses\ it is now straightforward to compute
the period matrix \tausw, and we find
\eqn\tauswsum{\tau_{mk} = { 2 i \over N} \sum_{n=1}^{N-1}
{f(v^N,n/N) \over f(-v^N,n/N)} \sin(\pi n/N) e^{i 2 \pi n(m-k)/N}.}
In obtaining this result we have multiplied the inverse $\tilde{a}$
matrix in \inverses\ on the right by the matrix $q^{-1}_{jk} = \delta_{j,k}
- \delta_{j-1,k}$ which converts the $\gamma$-period $\tilde{a}$
matrix to a $\beta$-period $a$ matrix.  While we cannot
simplify the period matrix any further for generic values
of $v$, at $v=0$ the sum is straightforward to do,
\eqn\tauzero{\tau_{mk} = {i \over N} (\cot(\pi (m-k+1/2)/N)
- \cot(\pi (m-k-1/2)/N)).}

There is a theorem in Riemann surface theory that says that
$\tau_{mk}$ is symmetric and that its imaginary part is
positive definite.  By replacing the summation index $n$ by
$N-n$ one easily shows that symmetry implies the condition that
\eqn\symmeqn{{f(v^N,n/N) \over f(-v^N,n/N)} = {f(v^N,1-n/N) \over
f(-v^N,1-n/N)}.}
For $v=0$ this condition holds trivially as both sides are one,
and moreover from the explicit expression given in \tauzero.
For generic values of $v$ we have checked numerically that
it does hold for various values of $v$ and $N$, which provides a check
on our explicit expression for the period matrix.

For positivity we again do not have a direct analytic proof that
this condition holds for the explicit form for $\tau$ given
above.  Even for $v=0$ we have not been able to show analytically
that all eigenvalues are positive.  Curiously though it is
straightforward to check that $\sin(2 \pi k j/N)$ for
$1 \leq j \leq [N/2]$ are eigenvectors of ${\rm Im}[\tau_{mk}]$
with eigenvalues $2 \sin(\pi j/N)$, which are indeed
positive.  For the remaining eigenvalues and for more
general values of $v$, we have checked numerically that
the eigenvalues are indeed positive.

\subsec{Probe in Enhancon Background}

In this section we compute the period matrix in various limits along a
one dimensional slice of moduli space corresponding to a wrapped
D7-brane probe of the enhancon background. For the most part we are
able to derive explicit expressions for finite $N$, and consider the
large $N$ expansion only when necessary.

The moduli in this case are given by
\eqn\probemod{\eqalign{\phi_j & = 
- {v \over N}, ~~~ 1 \leq j \leq
(N-1) \cr
\phi_{N} & = v - {v \over N}}}
where $v$ is the complex coordinate which parametrizes the
subspace that we will be investigating.
The associated curve ${\cal C}$ is given by
\eqn\probecurve{Y^2 = (X-v+v/N)^2 (X+v/N)^{2(N-1)} -
\Lambda^{2N}.}
\vskip 16pt

{\it $v/\Lambda \ll 1$ Limit}
\vskip 8pt

For $v/\Lambda \ll 1$ 
the branch points of the curve are
given by the series expansion in $v$
\eqn\branchpoints{X_k = \Lambda e^{i \pi k/N}\left(1 + {1 \over 2N}
({v \over \Lambda})^2 e^{-i 2 \pi k/N} \left(1 - {1 \over N}\right)
+ {\cal O}\left(({v \over \Lambda})^3\right)\right).}
This form of the branch points suggests a more convenient
basis for the holomorphic 1-forms than those given in \dlamdphi\ 
(which, as discussed above, are not independent when more than one of
the moduli space coordinates $\phi_k$ are equal).
Specifically we shall take the basis of holomorphic forms
given by
\eqn\newbasis{\omega_n = -{1 \over 2 \pi i} {X^{n-1} dX \over Y}, ~~~
1 \leq n \leq (N-1).}

The advantage of this basis is that the linear in $v$ correction
to the period matrix vanishes identically, as we now explain
in more detail.  We shall take the same basis of cycles
as in the previous example for $v=0$.  That is, the $\alpha_k$
cycle encircles the $X_{2k}$ and $X_{2k+1}$ branch points,
while the $\gamma_j$ cycle encircles the $X_{2j-1}$ and
$X_{2j}$ branch points.  The $\beta_k$ cycles are then 
defined as in \betacycles. 

To evaluate a period integral of any of the
holomorphic forms $\omega_n$ over an  $\alpha_k$ cycle, it
is convenient to 
parametrize the integral in terms of an angular variable $\varphi$ as
\eqn\intparam{X = \Lambda e^{i \pi 2k/N} e^{i \varphi/N}\left(1 + {1 \over 2N}
({v \over \Lambda})^2 e^{-i 4 \pi k/N} e^{-i 2 \varphi/N} (1 - {1 \over N})
+ {\cal O}\left(({v \over \Lambda})^3\right)\right),}
a result which follows naturally from the form of the
branch points \branchpoints.  Similarly for
a period integral about a $\gamma_k$ cycle, one simply replaces 
$k \rightarrow (k-1/2)$
in \intparam.  In this parametrization $X^{n-1} dX$ clearly
has no linear in $v$ piece.  Furthermore $Y$ has no linear
in $v$ piece as follows by substituting \intparam\
into \probecurve\
and expanding, or more simply by noting that $dX(v)/dv$ vanishes at
$v=0$ for $X(v)$ a branch point of the curve which
follows from differentiating \probecurve\
with respect to $v$.

The periods are now straightforward to write down in 
an expansion in $v$.  Specifically we find
\eqn\daDdtprobe{\eqalign{{\partial a_{Dm} \over \partial \xi_n} 
& = \oint_{\alpha_m}
\omega_n =
{1 \over 2 \pi N} \Lambda^{-(N-n)} \bigg(F(n) m_{mn} \cr
& +\left. ({v \over \Lambda})^2 {(N-1)(n-2) \over 2 N^2} F(n-2) m_{m,(n-2)} + 
{\cal O}\left(({v \over \Lambda})^3\right)\right), \cr
{\partial \at_{k} \over \partial \xi_n} & = \oint_{\gamma_k}
\omega_n = {\partial a_{Dm} \over \partial \xi_n}\bigg|_{m=k-1/2} }}
where we have defined
\eqn\intdef{\eqalign{F(n) & \equiv \sqrt{2} e^{-i \pi/4} 
\int_{0}^{\pi} d \varphi {e^{i(n/N -1/2) \varphi}
\over \sqrt{\sin(\varphi)}} \cr
& =-i4 \sqrt{\pi} {N \over n} {\Gamma(1+n/(2N)) \over \Gamma(1/2 +
n/(2N))}
e^{in \pi/(2N)} \sin(n \pi/(2N))}}
and the matrix
\eqn\matdef{m_{kn} \equiv e^{i 2 \pi k n/N}.}
The $\partial a_{k}/\partial \xi_n$ periods are integrals over 
$\beta_k$-cycles
defined in terms of the $\gamma$ cycles in \betacycles.

To compute the period matrix we need the inverse
\eqn\probeinv{\eqalign{\left( {\partial {\tilde a} 
\over \partial \xi}\right)^{-1}_{nj} & = 
-{i N \Lambda^{N-n} \over F(n)} e^{i \pi n/N} \bigg( (m^{-1})_{nj}
-({v \over \Lambda})^2 \sum_{l,k=1}^{N-1} (m^{-1})_{nl} \alpha_2(l,k)
m_{lk} (m^{-1})_{kj} \cr
& + {\cal O}\left( \left({v \over \Lambda} \right)^3\right) \bigg)  }}
where we have defined
\eqn\alphatwo{\alpha_2(k,n) \equiv {(N-1)(n-2) \over 2 N^2} {F(n-2) \over
F(n)}
e^{-i 4 \pi (k+1/2)/N}.}
The period matrix is then given by
\eqn\periodprobe{\eqalign{\tau_{mk} & = \tau^{double}_{mk}|_{v=0} + 
({v \over \Lambda})^2 \sum_{n,j=1}^{N-1} (\alpha_{D_2}(m,n) m_{mn}
(m^{-1})_{nj} \cr
& - \sum_{l,p=1}^{N-1} m_{mn} (m^{-1})_{nl} \alpha_2 (l,p)
m_{lp} (m^{-1})_{pj})(\delta_{j,k} - \delta_{j-1,k}) + {\cal O}
(({v \over \Lambda})^3)}}
where the first term is the period matrix of the double
enhancon evaluated at $v=0$ found in the previous subsection
and we have defined
\eqn\alphaDtwo{\alpha_{D_2}(m,n) \equiv {(N-1)(n-2) \over 2 N^2} {F(n-2) \over
F(n)}
e^{-i 4 \pi m/N}.}
The last two Kronecker delta functions convert the $\at$ periods
to $a$ periods as discussed above.

To connect this result with the supergravity description, 
we compute the induced metric on the
subspace of moduli space parametrized by $v$.  This
induced metric corresponds to the moduli space metric of a probe $D7$
brane in the enhancon background, which may be computed using the
Born-Infeld action plus Chern-Simons couplings.  To construct this
induced metric we need the periods $a_k$ as functions
of $v$.  These are given by the contour integrals \periods.
The computation is similar to what we have done before.
Namely one parametrizes the integrals as \intparam\ and
expands the integrand in powers of $v$.  The end result is
\eqn\aks{\eqalign{a_{k}(v) & = - {\Lambda \over i 4 \pi \sin(\pi/N) } 
\bigg(F(N+1) (1 - e^{i 2 \pi k/N}) + ({v \over \Lambda})^2 {(N-1) \over
2 N^2} F(N-1) (1 - e^{-i 2 \pi k/N}) \cr 
& +{\cal O} \left( \left({v \over \Lambda} \right)^3 \right)
\bigg).}}
The induced metric is now given by 
\eqn\induced{\eqalign{ds^{2}_{probe} & = \sum_{m,k=1}^{N-1}
{\rm Im}[\tau_{mk}] ({d a_m \over
dv})^* {d a_k \over dv} |dv|^2 \cr 
& = {1 \over 8 \pi^2 N} {|F(N-1)|^2 \over
\Lambda^2 \sin(\pi/N)} |v dv|^2 \cr
& = {1 \over 8 \pi^2 N} {|F(N-1)|^2 \over
\Lambda^2 \sin(\pi/N)} (d\rho^2 + 4 \rho^2 d\phi^2)}}
where the $(\rho,\phi)$ coordinates are defined as
$v \equiv \sqrt{2 \rho} \exp (i \phi)$.
The last form shows in particular that the induced metric
has a conical singularity at $\rho=0$ with negative deficit
angle.

One may use this result to compute the trajectory of a probe brane
approaching the enhancon shell. At the enhancon point, the probe hits
the conical
singularity in the induced metric. The full metric is smooth at this
point, and may be used to continue the motion of the probe past the
conical singularity. On the subspace parametrized by $v$, the
$\phi_i$ respect a $Z_{N-1}$ symmetry. Since the periods $a$ 
vary continuously (and with continuous first derivatives) through the
enhancon point, the same symmetry will be present after passing
through the enhancon point. The unique vacuum expectation value that
preserves this symmetry corresponds to the 
subspace parametrized by $v$, so the probe remains on the v-slice.
From the point of view of the spacetime
coordinates, the probe merges with the enhancon, and then smoothly reemerges
from the other side.

\vskip 16pt
{\it $v/\Lambda \gg 1$ Limit}
\vskip 8pt

In this section we consider the opposite extreme for
the probe, $v / \Lambda \gg 1$.  In this limit the probe
is located far from the remaining moduli as is clear
from the polynomial \probecurve.  In particular one
finds for the branch points in this limit 
\eqn\branchfar{\eqalign{\Xt_{\pm} & = {v \over \Lambda}\left(1
\pm ({\Lambda \over v})^N\right), \cr
\Xt_j & = x_j \left(1+ {1 \over N}{\Lambda \over v} x_j - {1 \over 2 N}
({\Lambda \over v})^2 (x_j)^2 + {\cal O}\left({ x_j \Lambda\over v }\right)^3\right)}}
where
\eqn\xjdefn{x_j \equiv ({\Lambda \over v})^{1/(N-1)} e^{i \pi (j-1)/(N-1)}}
for $1 \leq j \leq 2(N-1)$.  We have here introduced the scaled
and shifted variable $\Xt \equiv (X + v/N)/\Lambda$.

\ifig\ftwo{Branch cuts and 1-cycles for the probe configuration.}{\epsfxsize4.5in\epsfysize3.0in\epsfbox{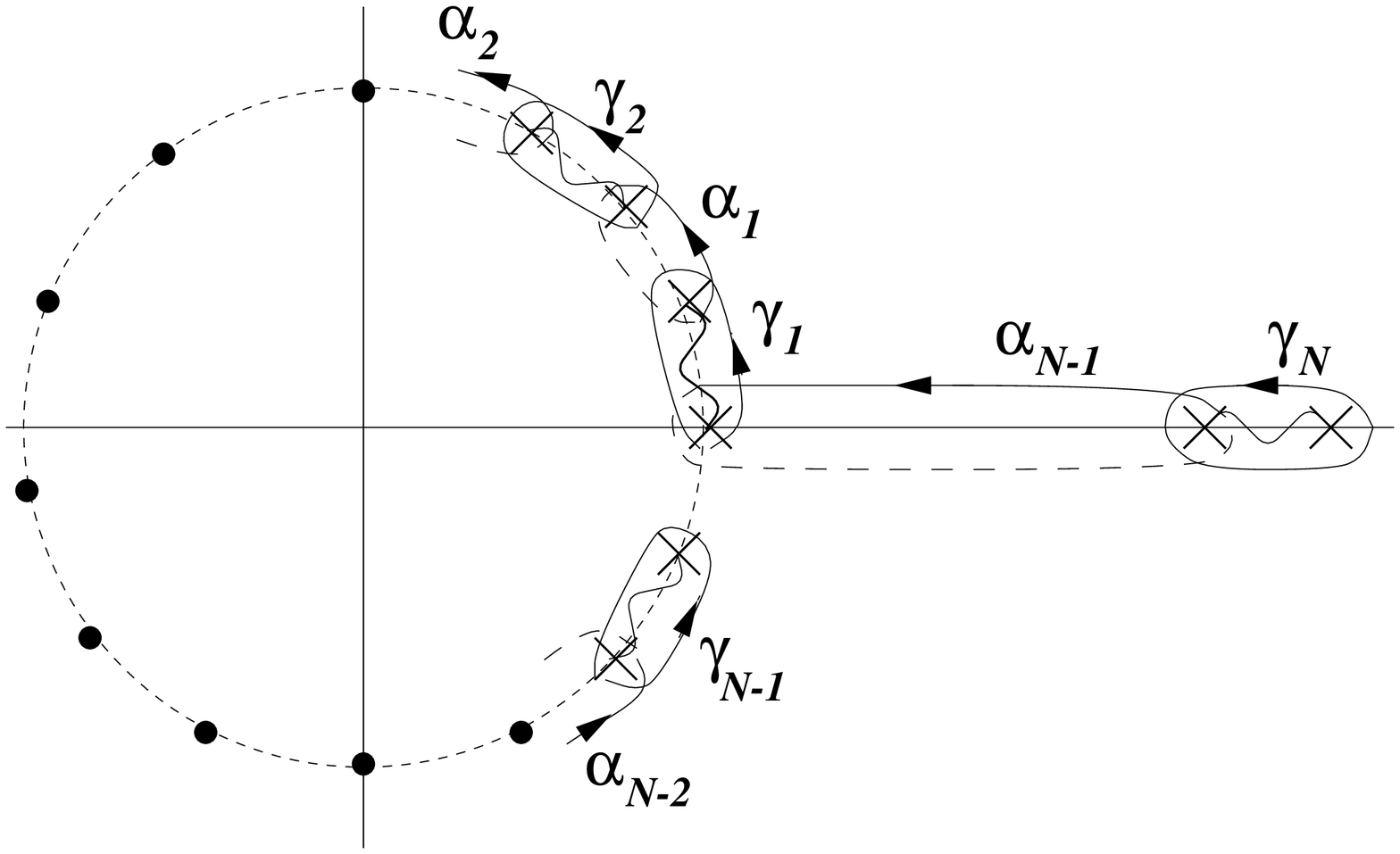}}

So far our choice of electric versus magnetic cycles
has been relatively arbitrary.  However in this case
we must be careful to identify the $\gamma_N$-cycle
which encircles the branch cut connecting the $\Xt_{\pm}$
branch points, see \ftwo, with an electric contour
and eg. $\alpha_{N-1}$, the cycle encircling the
$\Xt_{-}$ and $\Xt_1$ branch points, with a magnetic contour. 
For the remaining cycles
we take $\gamma_j$ to encircle $X_{2j-1}$ and $X_{2j}$
for $1 \leq j \leq (N-1)$
and $\alpha_k$ to encircle $X_{2k}$ and $X_{2k+1}$
for $1 \leq k \leq (N-2)$.
We further define $\beta$-cycles as
\eqn\betafar{\eqalign{\beta_j & \equiv \gamma_N +
\sum_{k=1}^{j} \gamma_k = - \sum_{k=j+1}^{N-1} \gamma_k, ~~
1 \leq j \leq (N-2) \cr
& \equiv \gamma_N =  -  \sum_{k=1}^{N-1} \gamma_k, ~~
j = (N-1).}}
The periods $a_j$ and $a_{Dm}$ are then defined as in \periods.

To compute the period matrix we again need the matrices of
integrals of some basis of holomorphic forms over the $\alpha$ and
$\beta$-cycles.  We shall take a slightly different basis for
the holomorphic forms than in the previous two cases, namely
let
\eqn\holofar{\omegat_n \equiv - {1 \over 2 \pi i}
{ \Xt^{n-1} d \Xt \over Y}, ~~ 1 \leq
n \leq (N-1).}
The only difference as compared to \newbasis\ is to
replace $X$ by $\Xt$, so clearly the basis \holofar\
is just a linear transformation of \newbasis.

 
The periods encircling pairs of branch points on
the circle are straightforward to compute in an expansion
in powers of $(\Lambda/v)$ and basically
follow from the previous two sections.
Specifically one finds
\eqn\periodsfar{\eqalign{{\partial \at_{j} \over \partial \xit_n} 
& = \oint_{\gamma_j}
\omegat_n =
- {1 \over 2 \pi (N-1) \Lambda^{N}} (x_{2j-1})^{n} (\Ft(n) 
+ {\cal O}({\Lambda \over v})), ~~ 1 \leq j,n \leq (N-1) \cr
{\partial a_{Dm} \over \partial \xit_n} & = \oint_{\alpha_m}
\omegat_n = {\partial \at_{k} \over \partial \xit_n}|_{k=m+1/2}, ~~
1 \leq n \leq (N-1), ~ 1 \leq m \leq (N-2)}}
where we have defined
\eqn\Ftdefn{\Ft(n) \equiv F(n)|_{N \rightarrow (N-1)}}
where $F(n)$ was given in \intdef.

Computation of the remaining periods over the $\alpha_{N-1}$
cycle is slightly more involved.  As it turns out we have
not been able to evaluate all of the remaining periods
directly.  However, as we discuss in the appendix, one
can evaluate them indirectly and still compute the
period matrix.  We record the results here and relegate
the details to the appendix.  For the remaining periods we find
\eqn\solveperiods{{\partial a_{D_{N-1}} \over \partial \xit_n}
\approx - {\Ft(n) \over 2 \pi (N-1) \Lambda^N} {e^{i \pi n \over N-1}
-1 \over e^{i2\pi n\over N-1}-1}
\left({v \over \Lambda} \right)^{-n  \over N-1}}
for $1 \leq n \leq (N-2)$ while the $n=(N-1)$
case is given by
\eqn\solvelast{
{\partial a_{D_{N-1}} \over \partial \xit_{N-1}} \approx 
i {(N+1) \over \pi \Lambda^N} {\Lambda \over v}
\log \left({v \over \Lambda} \right).}

The period matrix can now be expressed as
\eqn\tausw{\tau_{mj} =
\sum_{n,k=1}^{N-1} {d a_{D_m} \over d \xit_n}
 \left( {d \at \over d \xit}  \right)^{-1}_{nk} (\qt)^{-1}_{kj}.}
The $\qt_{jk}$ matrix was essentially defined
in \betafar\ as the matrix which converts
the $\gamma$-cycles (or $\at$ periods) into 
$\beta$-cycles (or $a$ periods), analogously to
the earlier discussion around \atotildea.
The inverse matrix does the reverse, and is given by 
\eqn\qinvfar{\eqalign{(q^{-1})_{kj} & \equiv (\delta_{k,j} - 
\delta_{k,j+1}), ~~ 1 \leq k \leq (N-1), ~ 1 \leq j \leq (N-2) \cr
& \equiv - \delta_{k,1}, ~~ 1 \leq k \leq (N-1), ~ j=(N-1).}}
The period matrix is now straightforward to evaluate and we find to
leading order at large $v$
\eqn\periodfar{\eqalign{\tau_{mj} & = {i \over (N-1)} 
\left(\cot(\pi {m-j+1/2 \over N-1})
- \cot(\pi {m-j-1/2 \over N-1}) \right), ~~ 1 \leq m,j \leq (N-2) \cr
& = {-i \over (N-1)} {e^{i \pi (m-1/2)/(N-1)} \over
 \sin(\pi (m-1/2)/(N-1))} , ~~  
1 \leq m \leq (N-2),~ j=(N-1) \cr
& =  {i \over \pi} (N+1)
\log \left({v \over \Lambda} \right),~~ m=(N-1),~j=(N-1).} }
The row $\tau_{(N-1),j}$ for $1 \leq j \leq (N-2)$ then
follows by the fact that the period matrix is symmetric.

The general form of this result was to be expected.  
Indeed, one might think that when $(v/\Lambda) \gg 1$
the probe is located far from the enhancon and the physics
of the probe reduces to that of a weakly coupled $SU(2)$
gauge theory, which is given by the tree-level plus one-loop
result. In the large $N$ limit, this accounts for the 
$\tau_{N-1,N-1}$ entry, which agrees
with the supergravity probe brane metric \jpp. However, already at
order $1/N$ we see a shift in the coefficient of the log term. The
one-loop beta function would give a coefficient of $N-1$, rather than
the $N+1$ that emerges from our detailed calculation. This arises from 
the response of the $N-1$ background branes to the position of the
probe, which can generate such corrections due to the close spacing
${\cal O}(\Lambda/N)$ of 
the eigenvalues of the associated moduli. 
We also see instanton
corrections in the other entries, which do not fall off as $v$ becomes
large, but are also suppressed by $1/N$ factors. These have a similar
interpretation.

\bigskip

{\it ${v \over \Lambda} \approx 1$ Limit}

We have so far discussed the large and small $v/\Lambda$
limits.  
Starting from $v/\Lambda \gg 1$ and decreasing
$v$, the probe branch points move toward the 
enhancon branch points distributed on (an approximately)
unit circle.  On the real axis there is a critical point
\eqn\crit{{v \over \Lambda} = {N^{1/N} \over (1 - 1/N)^{(1-1/N)}}
\approx 1 + {\log N \over N}}
at large $N$,
where one of the probe branch points collides with an enhancon
branch point.  At this point the period matrix becomes
singular, and a dyon becomes massless. For finite $N$, the singularity
is of the same form as the singularities of the $SU(2)$ theory \seiberg.
Near, but not at, this point however the period matrix is nonsingular.
To fully understand the motion of the probe in this region we must compute
the period matrix. We will obtain analytic expressions for the masses
of charged particles near this point, in a large $N$ limit, 
but so far have been unable to
obtain an analytic expression for the period matrix. 
Some of these periods have also been computed in \ferrari, and it has
been argued in \fstring, that the large $N$ limit can be taken in such 
a way as to
define a four-dimensional non-critical string theory.

The roots of the Seiberg-Witten curve for $v/\Lambda < 1$
 are given approximately in a large $N$ expansion by
\eqn\rootsnear{\Xt_k = e^{i \pi k/N} \left(1 - {1 \over N} \log (1 - 
{v \over \Lambda} e^{-i \pi k/N}) + -{1 \over 2}
({v \over N \Lambda} e^{- i \pi k/N})^2 + {\cal O}(1/N^3) \right)}
for $1 \leq k \leq 2N$ where $\Xt$ was defined after
\xjdefn. This expansion is valid provided the log term is much smaller 
than $N$. We keep $v/\Lambda$ fixed as $N$ becomes large.

To compute the periods we use the same set of holomorphic
one forms \holofar\ as given in the previous section.
Furthermore we shall take the same set of cycles as in the small $v/\Lambda$
case.  Parameterizing the integrals in terms of
an angular coordinate $\varphi$ as 
\eqn\paramnear{\Xt (\varphi) \equiv \Xt_{k} |_{k \rightarrow k +
\varphi/\pi},}
the expansion of the holomorphic one forms \holofar\
produces to leading order in $1/N$ the expression
\eqn\holoexp{\omegat_{n} = - {d \varphi \over 2 \pi N}
{e^{i \varphi n/N} \over \sqrt{e^{i2 \varphi} -1}}
{e^{i2 \pi kn/N} \over (1 - (v/\Lambda) e^{-i2 \pi k/N}
e^{-i \varphi/N})^{(n-1)/N}} + \cdots.}
Expanding the exponential $\exp(i \varphi/N)$ in powers
of $N$, one is left with integrals that we have seen
before, \intdef.  In particular one finds 
\eqn\dpercrit{\eqalign{{\partial a_{D_k} \over \partial \xit_{n}}
& = - {1 \over 2 \pi N} 
{e^{i 2 \pi k n/N} \over (1 -(v/\Lambda) e^{-i 2 \pi k/N})^{(n-1)/N}}
\bigg(1
 + 2 F(n) \bigg( {v \over \Lambda} {(n-1) \over N^2} {\log(1-
(v/\Lambda) e^{-i 2 \pi k/N}) \over (e^{i 2 \pi k/N} - v/\Lambda)} \cr
& - {1 \over N} \log(1- (v/\Lambda) e^{-i 2 \pi k/N})
-{v \over N \Lambda} {1 \over (e^{i 2 \pi k/N} - v/\Lambda)} \bigg)
+ 2 {v \over \Lambda} {n \over N} {(F(n) - F(n+1)) \over
(e^{i 2 \pi k/N} - v/\Lambda)} \cr
& + {\cal O} \left({1 \over N^2} \right) \bigg), \cr
{\partial \at_{m} \over \partial \xit_{n}} & = 
{\partial a_{D_k} \over \partial \xit_{n}}|_{k \rightarrow (m-1/2)
 }}.}
The periods are given in terms of the derivatives of the periods
by
\eqn\percrit{a_{D_k} = - N \Lambda \left(
{\partial a_{D_k} \over \partial \xi'_{N+1}} - {v \over \Lambda}
{\partial a_{D_k} \over \partial \xi'_{N}} + (1 - {1 \over N})
\left({v \over \Lambda} \right)^2 
{\partial a_{D_k} \over \partial \xi'_{N-1}}
\right) }
with similar expressions for the $a$'s and $\at$'s. As discussed in
\ferrari\ the large $N$ limit of the period $a_{D_{N-1}}$ is
non-uniform, 
in the sense that when $|v/\Lambda-1|$ 
decreases to of order $e^{-N}$, the large
$N$ expansion breaks down (this is because the log term that appears
in \rootsnear\ becomes of order $N$).

To compute the period matrix we must invert the matrix
$\partial a/\partial \xit$.  Unlike in our previous
cases however the $k$-dependence of $\partial a_k/\partial \xit_n$
is no longer contained only in a phase factor, but rather
something more complicated.  We do not know of any nice
analytic expression for the inverse of this matrix and
therefore have not been able to compute an analytic form
for the period matrix. Nevertheless the expressions for the periods
given above generate the masses of charged particles including
instanton corrections, at leading orders in the large $N$ expansion.

\newsec{Conclusions}

In this paper we have computed the moduli space metric for 
${\cal N}=2$ supersymmetric $SU(N)$ gauge theory along some distinguished
one-complex dimensional slices of relevance for resolution of
singularities in string theory via the enhancon mechanism. 
One
unexpected result of this analysis is the presence of corrections
suppressed by powers of $1/N$ that do not fall off at large
separations, in the regime where one of the classical
moduli is taken to be large.  When all moduli are well-separated,
one expects the tree-level plus one-loop contributions to only receive
corrections due to instantons, which
should be exponentially suppressed in
the large $N$ limit. The reason other corrections survive here is because the
moduli associated with the branes on the enhancon shell have
separations of order $1/N$, so we are never in a purely semiclassical
regime. 
In \ferrari, similar effects have been
attributed to fractional instantons.

From the supergravity perspective, this limit of moduli space
corresponds to a probe brane moving in the background of
$N-1$ branes on an enhancon shell. The corrections, which
are suppressed by powers of $1/N$,  contain
information about stringy and quantum corrections to the classical
supergravity description of the enhancon. 

We have also explicitly computed the enhancon metric near the point
where all the classical moduli vanish. As expected, 
the moduli space metric is smooth in
this limit. Gauge symmetry is not enhanced at this point -- the gauge
symmetry remains $U(1)^{N-1}$. The light particles include both
electrically and magnetically charged states which are mutually
non-local. The lightest masses go like $\Lambda/N$, which sets the
cutoff scale for our description of the low-energy effective field
theory.

We computed the induced metric on a one-complex dimensional
subspace corresponding to a probe brane near the enhancon locus, and
find a conical singularity at the center. In \refs{\johnson,\johnsontwo}
it was argued a
similar conical singularity appears on the induced metric on a
subspace of a higher dimensional analog of the Atiyah-Hitchin space,
relevant for a probe brane in the background of an enhancon shell 
arising from wrapped D6-branes.
As mentioned above, the full
moduli space metric at this point is smooth. This supports the picture
of a probe brane approaching the enhancon shell advocated in \jpp. The
probe brane melts into the enhancon branes and becomes
indistinguishable from them at the point where the classical moduli
vanish. In particular, a probe brane corresponding to a wrapped
D7-brane is unable to probe the region inside the enhancon shell.
We are currently completing a computation of the moduli space metric
for $SU(N)$ with fundamental matter. This will allow us to consider
instead a probe D3-brane which is able to probe inside the enhancon
shell. This will provide a useful check on the idea that the region
inside the enhancon shell is simply flat space.

Finally, it is interesting to address the question of to what extent
the enhancon mechanism is a general phenomenon. In the highly
symmetric situations we have studied in this paper, the presence of
the enhancon shell corresponds to the fact that the branch points of
the Riemann curve are not coincident, but rather become closely spaced
points around a circle. However nothing stops us considering other
regions in moduli space where branch points collide. In particular, we
could consider subspaces of moduli space where large numbers of branch
points collide and give rise to generalizations of the Argyres-Douglas
point \argyrdoug. Related multi-critical points have been studied at
large $N$ in \fstring.
These presumably give rise to timelike singularities in
the supergravity description for which the enhancon mechanism will not
work. We hope to return to a study of this regime in the future.

\bigskip
\centerline{\bf Acknowledgements}
\noindent
We thank F. Ferrari, A. Jevicki, R. Myers, J. Polchinski,
and S. Ramgoolam for helpful comments.
This research is supported in part by DOE
grant DE-FE0291ER40688-Task A.

\appendix{A}{Period matrix for $v/\Lambda \gg 1$}

We discuss the details of the computation of the
periods $\partial a_{D_{N-1}}/\partial \xit_n$ for
the $(v/\Lambda) \gg 1$ case in this appendix.
Some of these integrals
have essentially been done in \ferrari.  The basic point
is the following.  The integral is between the limits
$\Xt_1 \approx 1 + \log(\Lambda/v)/N$ and $\Xt_{-} \approx(v/\Lambda)(1-(\Lambda/v)^N)$.  
It follows that over most of the
integration region the $\Xt^{2(N-1)}$ term in $Y$ will dominate
so that the 1 can be dropped and one has the
simple integral
\eqn\approxzero{\eqalign{{\partial a_{D_{N-1}} \over \partial \xit_n}
& = \oint_{\alpha_{N-1}} \omegat_n \approx {1 \over i \pi
\Lambda^N} \int_{\Xt_1}^{\Xt_{-}} {d \Xt \over (\Xt - v/\Lambda)
\Xt^{N-n}} \cr
& \approx {i \over \pi \Lambda^N} {1 \over N-1-n} 
\left( {v \over \Lambda} \right)^{-n/(N-1)}, ~~ 1 \leq n \leq (N-2) \cr
& \approx i {(N+1) \over \pi \Lambda^N} {\Lambda \over v}
\log \left({v \over \Lambda} \right), ~~ n=N-1.}}
Actually the integral on the far right-hand-side of
the top line in \approxzero\ can be done exactly
and one would get ${\cal O}(N-n)$ terms.  We have written
only the dominant term above.  However for this approximation
to the full integral to be valid
one needs to check if the contribution of both the
other ${\cal O}(N-n)$ terms as well as the contribution
to the integral
coming from the integration region near the limits
where $Y$ vanishes is subdominant.  Actually
the former contribution can be removed easily
by taking a different basis of holomorphic one-forms.
For example, if we take the basis
\eqn\primebasis{\omega^{\prime}_n \equiv (\omegat_n -
{\Lambda \over v} \omegat_{n+1})}
for $1 \leq n < (N-2)$ and $\omega^{\prime}_{N-1} =
\omegat_{N-1}$ instead, the periods in the above approximation would
reduce to just one term.

The real problem is that the contribution to the periods
coming from the region of integration near the limits
becomes important
once $n = {\cal O}(1)$.  To see this consider the following correction
to the $\partial a_{D_{N-1}}/\partial \xi^{\prime}_n$ period
(we are working in the $\omega^{\prime}$ basis instead of
the $\omegat$ basis here simply because it decouples the
two effects mentioned above),
\eqn\correction{- {1 \over \pi i} \int_{\Xt_1}^{\Xt_m}
\left[ (1 - (\Lambda/v) \Xt) {\Xt^{n-1} \over Y} + {v \over \Lambda}
\Xt^{-N+n} \right] d\Xt,}
where $\Xt_1 < \Xt_m < \Xt_-$, but otherwise $\Xt_m$ is
arbitrary.  The integrand here is just the difference
of the exact holomorphic one-form \primebasis\ and
its approximate form given by dropping the 1 in
$Y$.  Expanding near the lower limit as
$\Xt = \Xt_1 (1 + x/(N-1))$ one obtains the integral
\eqn\correxp{\approx {1 \over \pi i} \left( {\Lambda \over
v} \right)^{n/(N-1)} {1 \over N-1} \int_{0}^{x_m}
\left[ {e^{x (n-1)/(N-1)} \over \sqrt{e^{2x} -1}}
- e^{-x (N-n)/(N-1)} \right] dx. }
The upper limit $x_m$ can be taken to infinity at the cost
of an exponentially small correction to the integral.
This integral can now be evaluated and one finds
\eqn\corrthree{- {1 \over \pi i}  \left( {\Lambda \over
v} \right)^{n/(N-1)} {1 \over N-1} \left[ - {N-1 \over N-n}
+ {\sqrt{\pi} \over 2} {\Gamma({N-n \over 2 (N-1)}) \over
\Gamma({2N -n -1 \over 2(N-1)}) } \right].}
This should be much less than the leading order
contribution to the period integral, which in this
case simply coincides with the contribution coming
from the first term in brackets of the above expression.
When $n=N-{\cal O}(1)$ then the term in brackets vanishes
to leading order and indeed this correction is subdominant.
However when $n= {\cal O}(1)$ the term in brackets does
not vanish to leading order and the ``correction''
term is of the same order as the leading order piece,
so the approximations made in this regime are not valid.

As we commented on in the text, we do not actually
need to compute all of these period integrals directly to
construct the period matrix.  We only
need to evaluate eg. 
$\partial a_{D_{N-1}}/ \partial \xit_{N-1}$
(we are working once again with the $\omegat$
basis) directly, for which the approximations
described above are valid, and the remaining
periods can be obtained indirectly as we now
discuss.  Basically one notes that given just
the periods in \periodsfar, and the
fact that the period matrix is symmetric, is enough to 
construct the entire period matrix except
for the element $\tau_{(N-1),(N-1)}$.  To compute
this element we need the values of the remaining
periods.  We can however solve for them by viewing
the symmetry requirement $\tau_{(N-1),k} = \tau_{k,(N-1)}$
for $1 \leq k \leq (N-2)$ as a set of equations
to determine the periods 
$\partial a_{D_{N-1}}/ \partial \xit_k$ over
the same range of $k$, where one must remember
that $\partial a_{D_{N-1}}/ \partial \xit_{N-1}$
has been given above.  In more detail, in terms of
the tilded period matrix defined as
$\tau_{mj} = \sum_k {\tilde \tau}_{mk} (q^{-1})_{kj}$,
the equations determining the remaining periods are
\eqn\remaining{\sum_{n=1}^{N-1} 
{\partial a_{D_{N-1}} \over \partial \xit_n}
\Big( \left( {\partial \at \over \partial \xit} 
\right)^{-1}_{nk}
- \left( {\partial \at \over \partial \xit} 
\right)^{-1}_{n,(k+1)} \Big)
= - {\tilde \tau}_{k,1}.}
It is a simple exercise to solve these equations for
the remaining periods to find \solveperiods.

\listrefs

\end